\documentclass[traditabstract]{aa} 
\usepackage{graphicx}
\usepackage{txfonts}
\usepackage{amssymb}
\usepackage[]{natbib}
\usepackage{longtable}
\usepackage{supertabular}

\usepackage[colorlinks=true, linkcolor=blue, citecolor=blue,
urlcolor=blue, breaklinks=true]{hyperref}
\PassOptionsToPackage{hyphens}{url}\usepackage{hyperref}

\newcommand{\tres}[0]{\mbox{TrES-2}}
\newcommand{\tresa}[0]{\mbox{TrES-2A}}
\newcommand{\tresb}[0]{\mbox{TrES-2b}}

\newcommand{\bjdtdb}[0]{\mbox{$\mathrm{BJD}_{\mathrm{TDB}}$}}
\newcommand{\bjdutc}[0]{\mbox{$\mathrm{BJD}_{\mathrm{UTC}}$}}
\newcommand{\hjdutc}[0]{\mbox{$\mathrm{HJD}_{\mathrm{UTC}}$}}

\begin{document}

\title{A consistent analysis of three years of ground- and space-based
photometry of TrES-2}
\author{S. Schr\"oter \and J.H.M.M. Schmitt \and H.M. M\"uller}
\institute{Hamburger Sternwarte, Universit\"at Hamburg, Gojenbergsweg 112, 21029
Hamburg, Germany\\ \email{sschroeter@hs.uni-hamburg.de}}
\date{Received November 29, 2011; accepted January 26, 2012}

\abstract { The G0V dwarf TrES-2A, which is transited by a hot Jupiter, is one
of the main short-cadence targets of the \textit{Kepler} telescope and,
therefore, among the photometrically best-studied planetary systems known today.
 Given the near-grazing geometry of the planetary orbit, TrES-2 offers an
outstanding opportunity to search for changes in its orbital geometry. Our study
focuses on the secular change in orbital inclination reported in previous
studies. We present a joint analysis of the first four quarters of
\textit{Kepler} photometry together with the publicly available ground-based
data obtained since the discovery of TrES-2b in 2006. We use a common approach
based on the latest information regarding the visual companion of TrES-2A and
stellar limb darkening to further refine the orbital parameters. We find that
the \textit{Kepler} observations rule out a secular inclination change of
previously claimed order as well as variations of the transit timing, however,
they also show slight indication for further variability in the inclination
which remains marginally significant.}
\maketitle

\section{Introduction}

The transiting Jovian planet \tresb \ was originally discovered by
\citet{ODonovan2006}, and later extensively observed by \citet{Holman2007} and
\citet{Sozzetti2007}, who determined more accurate properties for the \tres \
system. \tresa \ is an old ($\approx 5$~Gyr) G0V star with about solar radius
and mass, orbited by a close-in ($0.035$~AU) planet with a mass of $1.2$~$M_J$
and a period of $2.47$~d. The planet has a radius of $1.24$~$R_J$ and a large
inclination angle of about $84^\circ$, implying a nearly grazing transit. The
spectroscopic parameters of the systen determined by \citet{Sozzetti2007} were
confirmed and refined by \citet{AmmlervonEiff2009}.

As already pointed out by \citet{Holman2007}, \tres \ offers an outstanding
opportunity to search for changes in orbital parameters for several reasons: Its
large inclination and close-in orbit yield a high impact parameter, making the
shape and timing of the transit sensitive to variations of the orbital
parameters. Its large radius and short orbital period allow the study of large
numbers of deep transits over hundreds of epochs. Additionally, \tres \ is in
the field of view of the \textit{Kepler} telescope, which has already gathered
more than 220~days of publicly available photometric data, probably making \tres
\ the photometrically best-studied exoplanetary system to date.

Since its discovery there have been numerous attempts to detect secular changes
in the orbital parameters of \tresb \ due to an external perturber.
\citet{Mislis2009} combined their own transit observations with the data from
\citet{Holman2007} and found these data to be consistent with a decrease in
transit duration, suggesting a decrease of about $0.075^\circ/$yr in the orbital
inclination of the planet, an interpretation corroborated by another series of
observations \citep{Mislis2010}. Reverting to secular perturbation theory,
\citet{Mislis2010} suggest that a Jovian mass planet with a period between 50 to
100 days can explain the observed inclination changes. On the other hand,
\citet{Scuderi2010}, in their analysis from data taken in June 2009, find
systematically higher values for the inclination of \tresb, ruling out any
systematic trend. Their analysis also includes 33 days of \textit{Kepler} data
comprising multiple transits originally published by \citet{Gilliland2010},
which they find to be in agreement with their ground-based observations.

Additional photometric data of \tres \ are presented by \citet{Rabus2009}, who
analyze five transits sytematically searching for signatures of transit timing
variations. However, they do not find a clear signal and are only able to
provide upper constraints on the perturber's mass. \citet{Christiansen2011}
obtained nine additional transits from NASA's EPOXI Mission of Opportunity (Deep
impact). Comparing their results with earlier measurements, they find a formally
significant decrease in the transit duration with time, which however disappears
if they exclude the duration measurement by \citet{Holman2007}.

A thorough analysis of the first two quarters of \textit{Kepler} data (Q0 and
Q1) is presented by \cite{Kipping2011}, who do not find any inclination change
within the \textit{Kepler} data, obtaining an inclination change of
$(+0.0019\pm0.0020)^\circ/$cycle. They also reanalyze the data by
\citet{Holman2007} fitting the limb-darkening coefficients and compare the
measured average transit durations in both data sets. Their results neither
support nor exclude the inclination change, largely due to the still very small
temporal baseline of 18 cycles. \citet{Raetz2009, Raetz2011} analyze 22
additional transits observed between March 2007 and August 2010, but find no
indications of variations in the transit duration. However, their data quality
does not suffice to exclude the inclination change proposed by
\citet{Mislis2010}.

The whole discussion is further complicated by the results of
\citet{Daemgen2009}, who detected a visual companion to \tresa \ at a separation
of $1.089\pm0.008 \arcsec$ that contaminates the photometric data, and leads to
filter-dependent deviations in the orbital parameters if not being accounted
for. While \citet{Scuderi2010} already mention the contribution of a third light
and estimate its influence on the system parameters, \cite{Kipping2011} were
actually the first to include it in their light-curve modeling of the
\textit{Kepler} data.

The purpose of this work is to present a joint analysis of all publicly
available \textit{Kepler} data of \tres \ focussing on the apparent change in
orbital inclination as proposed by \citet{Mislis2010}. We exploit four quarters
of \textit{Kepler} photometry and publicly available ground-based transit
observations in a joint approach including the latest observational results
regarding the \tres-system, aiming at a consistent picture of the observational
status quo regarding changes in the orbital parameters, especially the
inclination of the planet.

\section{Light-curve modeling}

We fit the transit data with the transit model developed by
\citet{MandelAgol2002} making use of their \texttt{occultquad}
and \texttt{occultnl} FORTRAN
routines\footnote{\url{http://www.astro.washington.edu/users/agol}}. From the
transit light curve, we can directly infer the following parameters: the orbital
period, $P$, the time of transit minimum, $\tau$, the radius ratio, $p=R_p/R_s$,
the semi-major axis in stellar radii, $a/R_s$, the orbital inclination, $i$, and
the limb-darkening law. For our fits we either assume a quadratic limb darkening
with coefficients $u_1$ and $u_2$ or a four-parametric non-linear prescription
as proposed by \citet[][their Eq.~5]{Claret2011}, which provides a significantly
more accurate representation of model atmophere intensities
\citep{Howarth2011LD}. Furthermore, our model includes the contribution of a
third light, $L_3$, given as a fraction of the host-star's luminosity.

We assume a circular orbit fixing $e=0$, as there are no indications for any
non-zero eccentricity \citep{Kipping2011}.

\subsection{On the role of limb darkening}
\label{roleLD}

There is an ongoing debate on the role of limb darkening in the analysis of
planetary transit light curves regarding the question whether the limb-darkening
coefficients (LDCs) should be fitted together with the remaining parameters or
left fixed during the fit. On the one hand, LDCs are the only quantities in the
parameter set describing the transit that can be predicted theoretically, given
some basic spectroscopic information of the object. As such they are inherently
dependent on the quality of available spectroscopic data and the subtleties of
underlying assumptions entering the model calculations itself. Fitting the LDCs
therefore assures that the resulting transit parameters are independent of the
spectroscopic background model and that the uncertainties of the spectroscopic
parameters are readily included into the transit model. Furthermore, the
inclusion of LDCs as free parameters in the fitting process leads to
considerably larger uncertainties in the remaining model parameters, reflecting
the fact, that the LDCs are highly correlated to most parameters.

On the other hand, the data quality of most ground-based measurements is not
sufficient to substantially constrain the LDCs. Several fit parameters,
especially the LDCs, are known to be mutually correlated, as demonstrated
analytically by \citet{Pal2008} and numerically by \citet{Southworth2008} in the
specific case of \tres. The mutual correlations among the parameters are
particularly strong in the case of \tres, where the transit is nearly grazing,
because \tresb \ only covers the outermost parts of the stellar disk that are
dominated by strong limb-darkening effects. The remaining orbital parameters
thus depend crucially on the details of the adopted limb darkening resulting in
strong correlations with the LDCs. 
The amount of information on the limb darkening, which can be extracted
even from the \textit{Kepler} data, is limited by the fact, that the transit of
\tresb \ is nearly grazing, i.e.~$\mu \ll 1$. Assuming a quadratic limb-darkening
law this approximation yields
\begin{equation}
\frac{I(\mu)}{I_0}=1-(u_1+u_2)+\mu(u_1+2u_2)-\mathcal{O}(\mu^2).
\end{equation}
Therefore, to zeroth order we are only sensitive to the sum of the
LDCs, i.e. the amount of deducible knowledge on the individual parameter is
inherently limited.
This introduces a degeneracy between both LDCs due to their strong
correlation.
While, in principle, it would
always be preferable to fit for LDCs, thereby cross-checking spectroscopic
results and theoretical modeling, analyses of ground-based observations of \tres
\ cannot afford to refrain from including as much independently obtained
information as possible.

Therefore, in a first attempt, we will fit the \textit{Kepler} data without
assuming any a priori information on the LDCs (see Sect.~\ref{sec:fitKepler})
and compare our results to theoretical predictions. However, comparing the
results from the \textit{Kepler} data to several ground-based transit
observations requires a common treatment of the data. We therefore eventually
use fixed LDCs for all data sets assuming the theoretically predicted values
(see Sect~\ref{sec:fitAll}).

The spectroscopic studies by \citet{Sozzetti2007} and \citet{AmmlervonEiff2009}
concordantly suggest that \tresa \ has an effective temperature of 5800~K, a
surface gravity of 4.4, mictroturbulence velocity of 1~km/s and solar
metallicity (see Table~\ref{tab:SpecParm}). \citet{Claret2011} provide LDCs in
several filter bands calculated for a grid of PHOENIX and ATLAS9 model
atmospheres, employing two different calculation methods. We linearly
interpolated on the data grid to obtain a set of LDCs pertaining to a PHOENIX
model closely matching the inferred spectroscopic properties of \tresa.
Our interpolated
LDCs for the different filter bands are summarized in Table~\ref{tab:LDCs}.

\begin{table}
\caption[]{Spectroscopic parameters of TrES-2}
\label{tab:SpecParm}
\centering
\begin{tabular}{l l l l l}
\hline\hline
     & \citet{Sozzetti2007}  & \citet{AmmlervonEiff2009} \\
\hline
$T_\mathrm{eff}$ (K)  & $5850\pm50$    & $5795\pm73$   \\
$\log{g}$ (cgs)     & $4.4\pm0.1$    & $4.30\pm0.13$ \\
$\xi_T$ (km/s) & $1.00\pm0.05$  & $0.79\pm0.12$ \\
$[\mathrm{M}/\mathrm{H}]$        & $-0.15\pm0.10$ & $0.06\pm0.08$ \\
\hline
\end{tabular}
\end{table}

\begin{table}
\caption[]{Non-linear limb-darkening coefficients obtained by linear
interpolation on the tables provided by \cite{Claret2011} based on PHOENIX atmosphere 
models. The last column gives the fraction of flux due
to the visual companion detected by \citet{Daemgen2009}.}
\label{tab:LDCs}
\centering
\begin{tabular}{llllll}
\hline\hline
Filter & $a_1$ & $a_2$ & $a_3$ & $a_4$ & $L_3$ \\
\hline
\textit{Kepler} &  $0.473$\tablefootmark{a} & $0.192$\tablefootmark{a} &
- & - & 0.0276 \\ 
\textit{Kepler} &   0.801  & -0.676  & 1.053  & -0.388   & 0.0276 \\
Str\"omgren u   &   0.910  & -1.006  & 1.388  & -0.355   & 0.0045 \\
Str\"omgren v   &   0.863  & -0.767  & 1.872  & -0.617   & 0.0102 \\
Str\"omgren b   &   0.698  & -1.206  & 1.572  & -0.599   & 0.0180 \\
Str\"omgren y   &   0.758  & -0.663  & 1.203  & -0.454   & 0.0227 \\
Johnson U       &   1.085  & -1.412  & 1.721  & -0.466   & 0.0053 \\
Johnson B       &   0.754  & -0.937  & 1.734  & -0.641   & 0.0137 \\
Johnson V       &   0.742  & -0.602  & 1.134  & -0.434   & 0.0216 \\
Cousins R       &   0.796  & -0.632  & 0.993  & -0.378   & 0.0294 \\
Cousins I       &   0.853  & -0.746  & 0.963  & -0.363   & 0.0360 \\
SDSS i'         &   0.829  & -0.688  & 0.935  & -0.354   & 0.0343\tablefootmark{b} \\
SDSS z'         &   0.861  & -0.807  & 0.967  & -0.360   &
0.0425\tablefootmark{b} \\
\hline
\end{tabular}
\tablefoot{\tablefoottext{a}{Interpolated coefficients for a quadratic
limb-darkening law.} \tablefoottext{b}{Values are based on magnitude
differences directly measured by \citet{Daemgen2009}.}}
\end{table}

\subsection{On the contribution of a third light}

\citet{Daemgen2009} report the detection of a faint visual companion of \tresa \
in their ground-based high-resolution images obtained with the \textit{AstraLux}
lucky imaging camera at Calar Alto. The angular distance from \tresa \ to this
object is only $1.089\pm0.009 \arcsec$. It contaminates all previous published
observations and introduces a systematic error in the transit parameters,
especially the transit depth. According to \citet{Southworth2010}, roughly 5\,\%
of third light can be compensated for by decreasing $a/R_s$ by 1\,\%, $p$ by
3\,\% and $i$ by $0.1$~degrees.

\citet{Daemgen2009} measured magnitude differences in the Sloan Digital Sky
Survey (SDSS) z' and i' filters, that can be used to obtain the flux
contribution of the third light, $L_3$. They find $\Delta
\mathrm{z}'=3.429\pm0.010$ and $\Delta \mathrm{i}'=3.661\pm0.016$ resulting in
flux contributions of $0.0425\pm0.0004$ and  $0.0343\pm0.0005$ in the z' and i'
bands, respectively. We assume that the nearby companion has an effective
temperature of 4400~K, surface gravity of 4.5, solar metallicity and
microturbulent velocity of 2~km/s compatible with its spectral type being K4.5
to K6 as derived by \citet{Daemgen2009}. We used Gray's SPECTRUM together with a
Kurucz atmosphere model to calculate a synthetic spectrum. We then use STScI's
\textit{pysynphot}\footnote{\url{http://stsdas.stsci.edu/pysynphot/}} to
convolve the spectrum with transmission curves of all standard filter systems,
taking into account the high-resolution \textit{Kepler} response
function\footnote{\url{http://keplergo.arc.nasa.gov/}\newline
\url{CalibrationResponse.shtml}} to obtain the throughput in \textit{Kepler}'s
broad spectral band. The resulting fractional contributions of third light,
$L_3$, are tabulated in Table~\ref{tab:LDCs}.

\subsection{Fitting approach and error analysis}

To obtain adequate errors for the fit parameters and to determine the mutual
dependence of the parameters, we explore the parameter space by sampling from
the posterior probability distribution using a Markov-Chain Monte-Carlo (MCMC)
approach.

Typically, the system parameters, especially the LDCs, are prone to substantial
correlation effects. These correlations can easily render the sampling process
inefficient. It is therefore advisable to use a suitably decorrelated set of
LDCs as e.g. proposed by \citet{Pal2008}. However, due to the near-grazing
system geometry in the case of \tres \ we expect strong correlations between the
whole set of parameters. We therefore chose to use a modification of the usual
Metropolis-Hastings sampling algorithm, which is able to adapt to the strong
correlation structure. This modification, known as Adaptive Metropolis algorithm
\citep[AM;][]{Haario2001}, releases the strict Markov property of the sampling
chains by updating the sampling parameters using a multivariate jump
distribution whose covariance is tuned during sampling. This tuning is based on
all previous samples of the chain, so that AM looses the Markov property.
However, the algorithm can be shown to have the correct ergodic properties,
\mbox{i.e.} it approaches the correct posterior probability distribution under
very general assumptions \citep{Haario2001, Vihola2011}. We checked that AM
yields correct results for simulated data sets with parameters close to \tres,
and found that this approach showed fast convergence and efficiency.

All MCMC calculations make extensive use of routines of
\texttt{PyAstronomy}\footnote{\url{http://www.hs.uni-hamburg.de/DE/Ins/Per/Czesla/}\\ 
\url{PyA/PyA/index.html}}, a collection of Python routines providing a
convenient interface to fitting and sampling algorithms implemented in the PyMC
\citep{Patil2010} and SciPy \citep{Jones2001} packages.

\section{Data preparation}
\label{sec:dataprep}

We retrieved the \textit{Kepler} data of quarters Q0 to Q3 from the NASA
Multimission Archive at STScI\footnote{\url{http://archive.stsci.edu/kepler/}}.
Specifically, we use data pertaining to data release~5 (Q0, Q1), data release 7
(Q2), and data release~4 (Q3) as provided by the \textit{Kepler} Data Analysis
Working Group\footnote{\url{http://archive.stsci.edu/kepler/release_notes}}.
\textit{Kepler} observed \tres \ in short cadence mode with a sampling of
$58.85$~s covering 229~days. For our analysis, we use the raw aperture
photometry (SAP) flux and the corresponding error as provided in the FITS files.
The raw data have been processed using \textit{Kepler}'s Photometric Analysis
(PA) pipeline, which includes barycentric time correction, detection of
Argabrightening (a transient increase of background flux possibly due to dust
particles) and cosmic ray events, background removal, and the computation of
aperture corrected flux. Further, we removed invalid data points and all points
marked by the SAP\_QUALITY flag. The \textit{Kepler} data provide time stamps in
Barycentric Julian Dates (\bjdutc), which we convert into Barycentric Dynamical
Time (\bjdtdb) accounting for 34 leap seconds elapsed since 1961.

The data show discontinuities and exponential jumps due to instrumental duty
cycles (``pointing tweaks'', ``safe mode recoveries'', and ``earth points'') and
long-term trends (``focus drifts''), which must be corrected beforehand. First,
we removed all transits from the light curve including one half of the transit
duration on either side and rejected $3\sigma$ outliers using a sliding
median filter of window size 30 minutes. We then used the information provided
in the data release notes to exclude data points affected by safe mode
recoveries and earth points, which are difficult to correct otherwise.
Subsequently, we divided the data into chunks covering undisturbed
duty cycles. As there are two pointing tweaks occuring during quarter Q2, we
subdivided the Q2 data at these tweaks into three chunks covering full duty
cycles.

To remove long-term trends due to focus drifts, we followed an approach similar
to \citet{Kipping2011} by using the discrete cosine transform
\citep{Makhoul1980} to obtain a smoothed model of the continuum for each data
chunk separately. The discrete cosine transform decomposes a signal $x(n)$ into
a linear combination of $N$ cosine functions according to
\begin{equation}
y(k)=\sum_{n=0}^{N-1}x(n)\cos{\left(\frac{\piup k (2n+1)}{2N}\right)},
\hspace{5mm}0\leq k < N,
\end{equation}
and provides an efficient way of extracting the low-frequency information within
the signal. 
We applied the discrete cosine transform to each chunk of the light curve
removing all but the first $l$ low-frequency terms, where $l$ is the rounded
integer value of the length of the chunk divided by the orbital period. The thus
cleaned transformed signal was inversely transformed to obtain a smoothed model
of the continuum. Finally, we divided the complete light curve including the
transits by the model and extracted the data in a box of half-width $0.06$~days
around each transit center. We further discard three transits, which occur close
to the pointing tweaks, where the continuum flux shows strong distortions.

To check that the results of our study do not depend on the details of the data
reduction, we also extracted the un-detrended data directly around each transit
center and divided the transit by a third-order polynomial fit to the local
continuum \citep{Welsh2010}. We find, that this procedure yields essentially the
same results.

The ground-based data covering a total of 8 planetary transits between 2006 and
2009 are used as provided by the authors (see Table~\ref{tab:data}, for
references). The time stamps are specified as Heliocentric Julian Dates
(\hjdutc) and are converted into \bjdtdb \ using the webtool by
\citet{Eastman2010}.

\begin{table}[t]
\caption[]{Publicly available data of ground-based photometric observations of \tres. The epoch refers to the first transit of \tresb \ observed by \citet{ODonovan2006}.}
\label{tab:data}
\centering
\begin{tabular}{l l l}
\hline\hline
Epoch & Filter & Reference \\
\hline
13, 15, 34 &  z'      & \citet{Holman2007}  \\
263        &  near R  & \citet{Mislis2009}  \\
312        &  I       & \citet{Mislis2009}  \\
395        &  I       & \citet{Mislis2010}  \\
414        &  I,y,b,v & \citet{Mislis2010}  \\
421        &  I       & \citet{Scuderi2010} \\
\hline
\end{tabular}
\end{table}

\section{Fitting results}

Before fitting the \textit{Kepler} data, we use all available data to obtain
estimates of the period, $P$, and the time of transit minimum, $\tau$. As
already pointed out by \citet{Southworth2008}, $P$ and $\tau$ are usually
correlated with each other, but uncorrelated with the  remaining parameters. Therefore, the
period can be determined quite accurately without interference of other
parameters.  Fitting all publicly available data we obtained
$$ \tau_0=2453957.635486^{+0.000069}_{-0.000068}~\bjdtdb~\textrm{and} $$
$$P=2.470613402^{+0.000000150}_{-0.000000154}~\textrm{d,}$$
where $\tau_0$ refers to the first transit of \tresb \ observed by
\citet{ODonovan2006} and errors correspond to 68.3\,\% HPD (highest probability
density) intervals. The fractional error in $P$ is approximately $10^5$ times
smaller than the fractional errors in $a/R_s$ and $p$, and can safely be
ignored. We checked that including both parameters with Gaussian priors
corresponding to their errors in the MCMC calculations yields negligible changes
in the remaining parameters but slows down the MCMC algorithm. In the following,
we therefore fix these parameters to their best-fit values.

\subsection{Four quarters of \textit{Kepler} data}
\label{sec:fitKepler}

In the following, we fit all $81$~transits observed by \textit{Kepler} during
the first four quarters of observations. To explore the whole parameter space of
solutions, we relax the constraints on the parameters as much as possible by
imposing uniform prior probability distributions allowing large parameter
variations. By fixing the LDCs to their theoretically computed values we would
impose strong a priori information on the fitting process. If the LDCs were
chosen incorrectly, we would obtain well-constrained model parameters with
comparably small credibility intervals, which, however, could be inconsistent.
We therefore additionally fitted quadratic LDCs, assuming uniform priors on
$u_1$ and $u_2$, and performed an MCMC error analysis of the complete set of
parameters. Our best-fit solution is determined by those parameter values that
minimize the deviance after $10^6$~iterations of the sampler. The errors are
calculated from 68.3\,\% HPD intervals of the posterior probability
distributions for the parameters. Our results are summarized in the upper part
of Table~\ref{tab:results}.

Given their credibility intervals the parameters of our global fit are
compatible with those by \citet{Southworth2011} and \citet{Kipping2011}. We note
a slight discrepancy with the results by \citet{Kipping2011}, who obtain a
slightly larger radius ratio, $p$, of $0.1282$. A small, possibly systematic
discrepancy compared to the values of \citet{Kipping2011} has also been noted by
\citet{Southworth2011}. Within their $2\sigma$ error intervals our
LDCs are still consistent with the best-fit solution obtained by
\citet{Kipping2011}, who found LDCs of $u_1=0.52$ and $u_2=0.06$. 
We note that the sum, $u_1+u_2$, of the LDCs is very well constrained in
agreement with our considerations in Sec.~\ref{roleLD}.

\begin{figure}[t]
\centering
\includegraphics[width=0.5\textwidth]{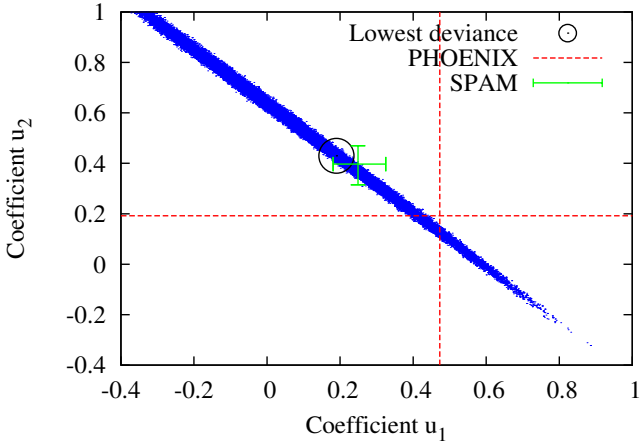}
\caption{Correlation between linear and quadratic limb-darkening coefficient resulting 
from $10^6$ MCMC runs. The lowest deviance combination of parameters (circle)
is significantly offset from the \citet{Claret2011} model prediction (dashed
red) but consistent with the result obtained from fitting synthetic
photometry based on the same atmosphere model (SPAM, green).}
\label{fig:ldcorr}
\end{figure}

As discussed in \citet{Howarth2011} we cannot directly compare our best-fit LDCs
to predictions based on the stellar atmosphere model. This is due to the fact
that we are fitting photometric data from a real star, using a simplified model
of the stellar limb darkening. The high impact parameter of \tresb \ together
with the fitting approach inevitably lead to considerable deviations of the
photometrically determined LDCs from model-atmosphere characterizations.
Instead, \citet{Howarth2011} proposes to compare the results from the
photometric analysis with the results obtained from \textit{s}ynthetic
\textit{p}hotometry based on a given
\textit{a}tmosphere \textit{m}odel (termed "SPAM"). Figure~\ref{fig:ldcorr}
shows the result of this consideration. The lowest deviance combination of LDCs
after $10^6$ MCMC samples is obviously inconsistent if compared directly to the
predictions by \citet{Claret2011} based on PHOENIX model atmospheres for \tres.
However, the SPAM model based on the same model atmosphere is consistent with
the best-fit LDCs. This strengthens our confidence in the reliability of the
theoretically calculated LDCs.

Note that the simplified prescription of limb darkening used in our
model also introduces small, but non-negligible systematic errors in the
remaining light curve parameters due to the strong correlations between all
parameters. In the following we therefore use the four-parametric, non-linear
limb-darkening law proposed by \citet{Claret2011}, holding the LDCs
fixed at their theoretically expected values to obtain the remaining transit
parameters. This, of course, substantially reduces the errors on our parameter
estimates because the strong correlations mediated by the limb-darkening
treatment diminuish. Our resuls after $10^6$ iterations of the MCMC sampler are
summarized in the lower part of Table~\ref{tab:results}.

\begin{table}[t]
\caption[]{Lowest-deviance parameters for the \textit{Kepler} data after $10^6$
MCMC samples compared to the results with fixed limb-darkening coefficients. 
Both fits assume a third light contribution according to Table~\ref{tab:LDCs}.}
\label{tab:results}
\centering
\begin{tabular}{lll}
\hline\hline
Parameter & Value & Error (68.3\,\% HPD) \\
\hline
$i$ ($^{\circ}$)   & $83.874$ & [ $83.806$ , $83.910$ ] \\
$p$ & $0.1257$ & [ $0.1240$ , $0.1265$ ] \\
$a/R_s$ & $7.903$ & [ $7.835$ , $7.937$ ] \\
$u_1$ & $0.19$ & [ $-0.10$ , $0.36$ ] \\
$u_2$ & $0.43$ & [ $0.25$ , $0.75$ ]\\
$u_1+u_2$ & $0.626$ & [$0.605$, $0.648$]\\
\hline
$i$ ($^{\circ}$) & $83.9788$ & [ $83.9740$ , $83.9837$ ] \\
$p$     & $0.12759$ & [ $0.12755$ , $0.12762$ ] \\
$a/R_s$ & $7.9899$  & [ $7.9856$ , $7.9943$ ] \\
$a_1$, $a_2$, $a_3$, $a_4$  & \multicolumn{2}{l}{fixed, see
Table~\ref{tab:LDCs}} \\
\hline
\end{tabular}
\end{table}

\subsection{Secular change in inclination}
\label{sec:fitAll}

\begin{table}[t]
\begin{minipage}[h]{0.5\textwidth}
  \caption{Prior information used for the MCMC calculations in
  Sec.~\ref{sec:fitAll}.}
  \label{tab:prior}
  \begin{center}
  \begin{tabular}[h]{l l l} \hline \hline
   Quantity & Prior & Source
   \\ \hline
   $a/R_s$  & $7.9899 \pm 0.0043$ &  Table~\ref{tab:results} \\
   $p$ & $0.127585 \pm 0.000035$ &  Table~\ref{tab:results} \\
   $\tau$ & uniform &  \\
   $i$ & uniform & \\
   $a_1$, $a_2$, $a_3$, $a_4$ & fixed & Table~\ref{tab:LDCs} \\
   $L_3$ & fixed & Table~\ref{tab:LDCs} \\
   $e$ & 0 (fixed) &  \\ 
   \hline 
  \end{tabular}
  \end{center}
\end{minipage}
\end{table}

The values inferred from the fit to the \textit{Kepler} data are the most robust
estimates of the planetary parameters of \tres \ currently available. In the
following, we reanalyze all publicly available photometric data of \tres \
examining each transit separately to search for secular variations of the
orbital inclination.

Secular inclination changes are empirically detectable as changes in transit
duration, which is a function of $a/R_s$, $p$ and $i$ \citep{Seager2003}. While
each of these parameters may itself exhibit secular changes resulting in changes
in the transit duration \citep{Kipping2010}, we focus in the following on
changes in the orbital inclination, $i$, and the time of transit minimum,
$\tau$, of each transit. We thereby assume that neither the radius ratio nor the
planetary orbit change in size over the course of the observations and that the
orbit is circular. However, we still must account for our incomplete knowledge
of the true value of $a/R_s$ and $p$. We therefore leave both parameters free
during the MCMC calculations imposing Gaussian priors according to the errors
obtained from our global fit. Depending on the instrumental setup of the
observations (cf. Table~\ref{tab:data}) we take into account the expected third
light contribution and the LDCs according to Table~\ref{tab:LDCs}. For each
transit we obtain the posterior distributions for both $\tau$ and $i$ using
$10^4$ iterations of the MCMC sampler. We summarize the prior information in
Table~\ref{tab:prior}. The errors again correspond to 68.3\,\% HPD intervals.

Concerning the \textit{Kepler} data, \citet{Kipping2011} did not find evidence
for parameter changes within the first two quarters. However, the public release
of the last two quarters, i.e. quarters Q2 and Q3, provided a substantial
enlargement of the data set. We therefore reanalyze the complete Q0 to Q3 set of
transits in detail searching for variations of the inclination or transit timing
variations (TTVs).

\begin{figure}[t]
\centering
\includegraphics[width=0.5\textwidth]{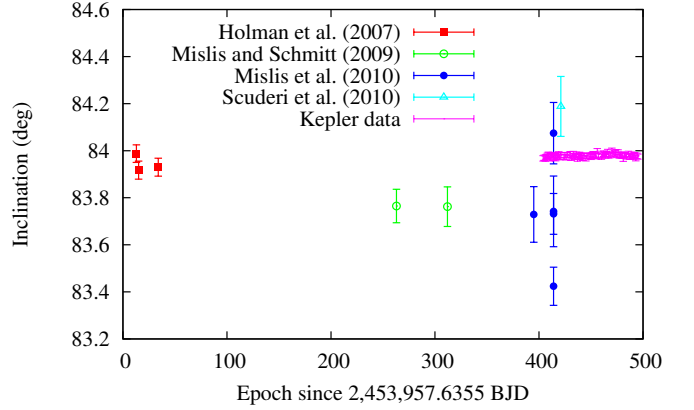}
\caption{Inclination versus transit epoch for \tres \ based on publicly
available ground-based data and the \textit{Kepler} observations. Errors
are derived from 95.5\,\% HPD intervals corresponding to $2\sigma$ errors based
on $10^4$ MCMC iterations for each transit.}
\label{fig:inclChange}
\end{figure}

Our results are shown in Fig.~\ref{fig:inclChange}. The superior quality of the
\textit{Kepler} data manifests itself in very small credibility intervals for
these observations. While the data by \citet{Holman2007} are fully consistent
with the inclination obtained by \textit{Kepler}, we find systematically lower
inclinations for the data sets of \citet{Mislis2009} and \citet{Mislis2010}. In
fact, ignoring the \textit{Kepler} data, the secular inclination change detected
by \citet{Mislis2009} is clearly visible.

Note, however, that our common fitting approach based on the results from
\textit{Kepler} and including the third light contribution yields best-fit
values for the inclinations derived from the ground-based measurements that are
closer to the \textit{Kepler} values than those originally obtained by
\citet{Mislis2009} and \citet{Mislis2010}. Furthermore, our new errors are quite
large so that the \textit{Kepler} value lies only just outside of our $3\sigma$
error.

We also checked, whether it is possible to completely reconcile the
\textit{Kepler} observations with the results by \citet{Mislis2009} and
\citet{Mislis2010}. Accounting for the possibility of normalization problems
during data reduction we introduce an additional normalization constant to our
model. Including this constant as another parameter of our MCMC marginally
enlarges the uncertainties on the inclinations. We thus do not expect the
deviation to stem from simple normalization problems.

Another cause for erroneous results would result from incorrectly taking into
account the contribution of the contaminating third light. To fully
reconcile the inclinations by \citet{Mislis2009} and \citet{Mislis2010} with the
best-fit inclination from the \textit{Kepler} data would require contributions
of a third light of 10~to 20\,\% of the total flux, which can be excluded from
the observations by \citet{Daemgen2009}. However, assuming a uniform prior for the
third light contribution and demanding it to be smaller than 10\,\%, the 
credibility intervals become substantially larger including the \textit{Kepler}
result. We thus conclude, that problems with the flux calibration possibly
caused by the reference star are the most likely cause for the observed
deviations.

In either case, the Kepler data rule out a decrease in inclination of the size
expected by \citet{Mislis2010}. The exact reason for their systematic deviation
remains unclear.
The \citet{Holman2007} data are consistent with the first four quarters of
\textit{Kepler} data. Note, however, that despite the long available time
baseline the errors of the \citet{Holman2007} data points are too large to draw
significant conclusions on the presence of other, smaller linear trends.

\subsection{Variability in the \textit{Kepler} data}

\begin{figure}[t]
\centering
\includegraphics[width=0.48\textwidth]{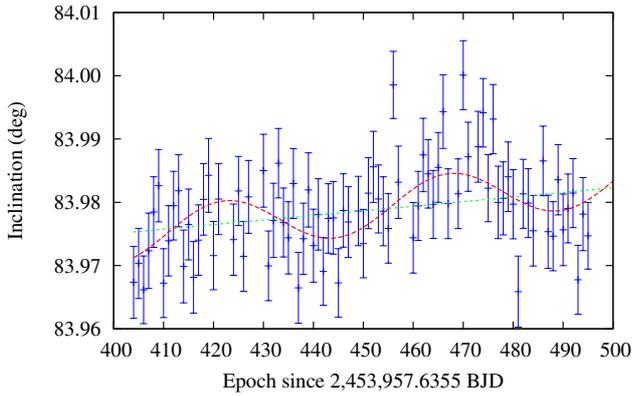}
\caption{Inclination versus transit epoch for the \textit{Kepler} observations.
Errors are computed from the 68.3\,\% HPD intervals based on $10^4$ MCMC
iterations for each transit. The best-fitting linear (dotted/green) and
sinusoidal models including a linear trend (dashed/red) are overplotted.}
\label{fig:inclChangeKepler}
\end{figure}

We now focus on the Kepler data searching for small trends or variations,
which could still be consistent with the previous ground-based observations.
Fig.~\ref{fig:inclChangeKepler} shows an enlarged view of the Kepler data; the
errors are derived from the 68.3\,\% HPD intervals, so they are
equivalent to $1\sigma$ errors. 

We found a slight linear trend in the \textit{Kepler} data. An MCMC analysis
yields a slightly non-zero slope of
\mbox{$(0.8\pm0.2)\times10^{-4}$~$^{\circ}$/cycle}. This would correspond to a
secular inclination change nearly two orders of magnitude smaller than the value
by \citet{Mislis2010}. Extrapolating this trend back to the epoch of transit
observations obtained by \citet{Holman2007}, we would expect an inclination of
$83.94^{\circ}\pm0.01^{\circ}$ consistent with the measurement errors.

In order to quantify the improvement of the different fits, we calculated the
statistic
\begin{equation}
  \hat{F}=\frac{(\chi_a^2-\chi_b^2)/(\nu_a-\nu_b)}{\chi_b^2/\nu_b}
\end{equation}
and compared its value to an $F$-distribution with $\nu_a-\nu_b$ and $\nu_b$
degrees of freedom \citep[e.g.,][]{Rawlings1998}.
We obtain a $p$-value of $0.004$ indicating that the linear trend provides a
better description compared to the constant at the $2.8\sigma$ level, which we
consider as indicative, but not significant evidence for the presence of
such a trend.

After subtracting the linear trend from the data, we computed an error-weighted
Lomb-Scargle periodogram \citep{Zechmeister2009}, which displays a single signal
at a period of 45 cycles or 111~days with a false-alarm probability below
10\,\%. Fig.~\ref{fig:inclChangeKepler} shows our best-fit sinusoidal model
including a linear trend. An $F$-test comparing the model to the constant model
yields a $p$-value of $0.015$ (or $2.4\sigma$); the test thus indicates the
presence of an additional sinusoidal modulation superimposed on the linear
trend, which however remains insignificant.

We can also compare the three models using the Bayesian information criterion,
$\mathrm{BIC}=\chi^2+k\ln{N}$ \citep{Schwarz1978}, which penalizes the number
$k$ of model parameters given $N=81$ data points. We obtain BICs of $136.3$,
$127.6$, and $117.0$ for the constant, the linear trend, and the sinusoid with
linear trend, respectively.

\begin{figure}[t] \centering
\includegraphics[width=0.48\textwidth]{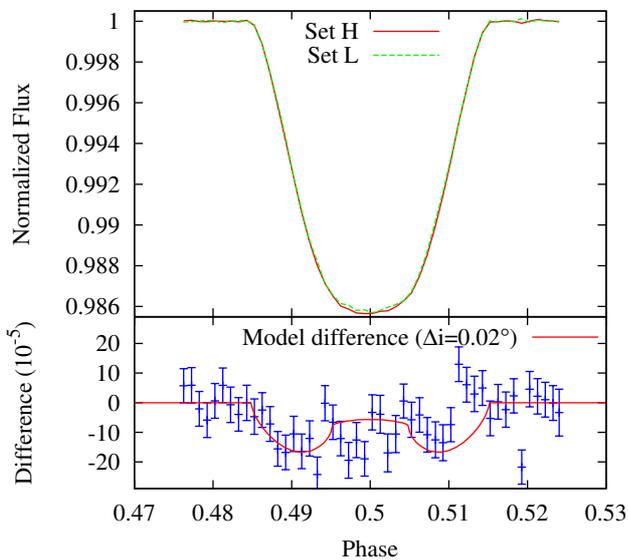} \caption{\textit{Upper
panel:} Averaged templates of two sets of $10$~transits with high (H, solid red)
and low (L, dashed green) inclinations, respectively, as expected from the
tentative sinusoidal variation of the inclinations obtained by \textit{Kepler}.
\textit{Lower panel:} The difference between the two averaged transits. The
expected difference for two models pertaining to $i=83.97^\circ$ and
$83.99^\circ$ is shown for reference (solid red).}
\label{fig:inclDiff}
\end{figure}

Despite being close to formal significance the signal is easily obliterated by
the noise and could be introduced by some subtleties in our analysis. Assuming
the sinusoidal model with linear trend to be true, we averaged two sets of
$10$~transits pertaining to low (around epoch~409, compare
Fig.~\ref{fig:inclChangeKepler}) and high (around epoch~470) inclinations,
respectively. Subtracting both templates, their difference should exhibit a
characteristic signature, which we model by subtracting two model calculations
for $i_L=83.97^\circ$ and $i_H=83.99^\circ$. We show the averaged templates
together with their difference and the model in Fig.~\ref{fig:inclDiff}. While
during ingress the flux difference shows a decrease following the difference
model, the signature during egress is obliterated by the noise. We checked that
averaging transits pertaining to epochs close to the zero-crossing points of the
sinusoidal model yields no substantial deviation from a constant.

We repeated our analysis using simulated control data to check, whether the
signal is introduced by our data reduction or by ``phasing effects'' due to the
\textit{Kepler} cadence \citep{Kipping2011a}. However, we could not detect
similar signals exceeding the noise. We further changed our data normalization
strategy (see Sec.~\ref{sec:dataprep}), but found the variation of inclination
to be stable against these changes. Analyzing similar \textit{Kepler} targets we
found spurious periodicities in some cases, which, however, remained
insignificant in the periodogram (at false-alarm probabilities above 10\,\%). In
summary, there seem to be some indications for a short-term inclination
variation and an underlying secular change in the \textit{Kepler} data of \tres.
However, the detected signal is merely an indication of inclination changes in
the \textit{Kepler} data, because in the baseline of space-based observations is
still too short to draw significant conclusions on the stability or variability
in the inclination (compare Fig.~\ref{fig:inclDiff}).

\begin{figure}[t]
\centering
\includegraphics[width=0.48\textwidth]{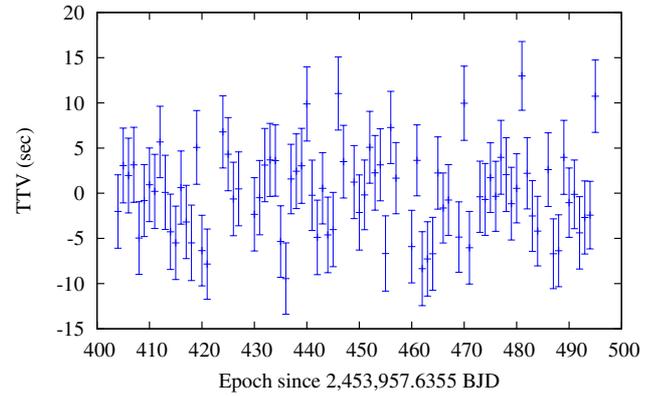}
\caption{Transit timing variation (TTV) versus transit epoch for \tresb \ using
the best-fit ephemerides. Errors are computed from the 68.3\,\%
HPD intervals based on $10^4$ MCMC iterations for each transit.}
\label{fig:T0ChangeKepler}
\end{figure}

Turning to the times of transit minimum of the Kepler transits,
Fig.~\ref{fig:T0ChangeKepler} shows that the transit timing is consistent with
being constant over the course of the \textit{Kepler} data. An $F$-test
comparing the constant and a linear model yields a $p$-value of $0.92$; thus,
there is no evidence for a linear trend. In the periodogram there is no signal
at false-alarm probabilities below 10\,\%. We thus do not see any evidence for
TTVs in the \textit{Kepler} data.

The apparent secular change in inclination refers to the orientation of the
orbital plane of \tresb \ relative to the observer's tangential plane and,
assuming a circular orbit, can be interpreted as nodal regression at fixed orbit
inclination. While gravitational perturbation due to an oblate host star is
unlikely given that \tresa \ is a slow rotator very similar to the Sun
\citep{Mislis2010}, additional bodies in the stellar system could offer a viable
explanation for secular changes in the orbital parameters. If we assume our
linear trend in inclination to be physical, we can estimate the period of a
perturber of specified mass in a circular, coplanar orbit using Eq.~13 in
\citet{Mislis2010}. We find that a perturbing planet of Jovian mass would be
expected at orbital periods between 100 and 300 days. However, a perturbing
second planet capable of causing nodal precession of the transiting planet would
also be expected to induce short-term periodic variations of its orbital
elements \citep{Holman2005, Agol2005}. To calculate the amplitude of such TTVs,
we use an approach based on analytic perturbation theory developed by
\citet{Nesvorny2008}. Outside possible mean-motion resonances additional Jovian
mass bodies with periods larger than 100~days would cause a TTV amplitude below
1~s if on a circular orbit. Given the standard deviation of the times of transit
minimum of 4.8~s, such a TTV amplitude cannot be ruled out. Thus, the existence of an
external perturber in the system still provides an entirely plausible
scenario.

\section{Summary and Conclusion}

We presented an analysis of all publicly available photometric data obtained
during planetary transits of \tresb \ including four quarters of data obtained
with the \textit{Kepler} space telescope. The transits were fitted in a common
scheme using theoretically calculated LDCs and taking into account the
contribution of a third light. 

Fitting quarters Q0 to Q3 of the \textit{Kepler} data, we were able to refine
the system parameters determined by \citet{Kipping2011} and
\citet{Southworth2011}. Our results are compatible with these previous studies,
our errors being slightly smaller. Its high impact parameter makes the
\tres-system sensitive to details of the adopted stellar limb-darkening model
because only the covered regions close to the stellar limb are used to
extrapolate the limb darkening for the whole stellar disk. Small differences
between limb-darkening prescriptions therefore yield significant differences in
the resulting best fits. We find that the theoretically calculated LDCs by
\citet{Claret2011} provide a correct description of the observed limb darkening
if compared with the data via the SPAM approach as advocated by
\citet{Howarth2011}.

The data by \citet{Mislis2009} and \citet{Mislis2010} are found to be
inconsistent with the remaing data, especially the \textit{Kepler} data. The discrepany can
neither be attributed to errors in the theoretical LDCs nor to the
contribution of third light, which the original studies did not account for.
The most probable explanations would be subtleties in the data reduction,
perhaps due to problems with the reference star, or the presence
of time-correlated noise yielding underestimated parameter errors
\citep{Carter2009}.

Excluding the \citet{Mislis2009} and \citet{Mislis2010} data sets, all
observations can be reconciled in a consistent picture. The extension of the
available \textit{Kepler} data by two additional quarters reveal some hints for
systematic variations in the inclination, the marginally significant slope would
be consistent with the data by \citet{Holman2007}. However, given the available
baseline of space-based observations we consider the inclination change as
indicative but not significant. Further data, which already have been obtained
by the \textit{Kepler} space telescope, will settle this issue in the near future.

Our study shows that space-based and ground-based photometry of \tres \ can be
analyzed in a common scheme yielding consistent results. While space-based
photometry offers the possibility to constrain all system parameters at once,
with the eventual decommissioning of \textit{Kepler} it will again be up to
ground-based measurements to search for long-term changes in the orbital
parameters of \tres.

\begin{acknowledgements}
This paper includes data collected by the Kepler mission. Funding for the Kepler
mission is provided by the NASA Science Mission directorate.
The Kepler observations were obtained from
the Multimission Archive at the Space Telescope Science Institute (MAST).
H.M.M. is supported in the framework of the DFG-funded Research Training Group
``Extrasolar Planets and their Host Stars"
(DFG 1351/2).
S.S. acknowledges support from the DLR under grant 50OR0703.
\end{acknowledgements}

\bibliographystyle{aa}
\bibliography{tres2Incl}

\Online

\begin{appendix}

\section{Results from individual fits}
In Table~A.1, we show the transit parameters of \tresb \ from individual fits of
ground-based and \textit{Kepler} data.

\renewcommand{\arraystretch}{1.22}
\longtab{1}{
\begin{longtable}{llllll}
\caption{Transit parameters for \tresb \ from individual fits of ground-based
and \textit{Kepler} data. Epoch refers to the first transit of \tresb \
observed by \citet{ODonovan2006}. For the MCMC analysis, we fix the
limb-darkening coefficients corresponding to the filter band (cf.~Table~\ref{tab:LDCs}) and use prior information as described in
Table~\ref{tab:prior}.}\\ \hline \hline Epoch & $\tau$ & $i$ & $a/R_s$ & $p$ & Reference\\ & [$\bjdtdb-2,450,000$~d] & [deg] &  &  & \\
\hline
\endfirsthead
\caption{continued.}\\
\hline \hline
Epoch & $\tau$ & $i$ & $a/R_s$ & $p$ & Reference \\
& [$\bjdtdb-2,450,000$~d] & [deg] &  &  & \\
\hline
\endhead
\hline
\endfoot
\endlastfoot
\multicolumn{6}{c}{Publicly available ground-based data} \\
\hline
$  13$ & $3989.752878_{-0.000233}^{+0.000208}$ &  $83.9876_{-0.0201}^{+0.0171}$
&  $7.9899_{-0.0047}^{+0.0038}$ &  $0.127585_{-0.000035}^{+0.000034}$ & \citet{Holman2007} \\ 
$ 15$ & $3994.694003_{-0.000227}^{+0.000233}$ & 
$83.9174_{-0.0219}^{+0.0166}$ &  $7.9898_{-0.0044}^{+0.0042}$ & 
$0.127585_{-0.000033}^{+0.000038}$ & \citet{Holman2007} \\ 
$  34$ & $4041.635919_{-0.000255}^{+0.000213}$ &  $83.9300_{-0.0198}^{+0.0184}$
&  $7.9893_{-0.0041}^{+0.0042}$ &  $0.127582_{-0.000031}^{+0.000039}$ &
\citet{Holman2007} \\ 
$ 263$ & $4607.398143_{-0.000574}^{+0.000512}$ & 
$83.7650_{-0.0332}^{+0.0378}$ &  $7.9897_{-0.0043}^{+0.0042}$ & 
$0.127585_{-0.000034}^{+0.000037}$ & \citet{Mislis2009} \\ 
$ 312$ & $4728.465365_{-0.000580}^{+0.000596}$ &  $83.7622_{-0.0413}^{+0.0430}$
&  $7.9895_{-0.0042}^{+0.0040}$ &  $0.127581_{-0.000034}^{+0.000036}$ &
\citet{Mislis2009} \\ 
$ 395$ & $4933.527211_{-0.000861}^{+0.000743}$ & $83.7291_{-0.0579}^{+0.0602}$ &
$7.9897_{-0.0045}^{+0.0039}$ &  $0.127584_{-0.000034}^{+0.000037}$ &
\citet{Mislis2010} \\ 
$ 414$ & $4980.466843_{-0.000717}^{+0.000712}$ & 
$83.4239_{-0.0358}^{+0.0455}$ &  $7.9897_{-0.0046}^{+0.0040}$ & 
$0.127583_{-0.000036}^{+0.000034}$ & \citet{Mislis2010} \\ 
$ 414$ & $4980.467292_{-0.000648}^{+0.000552}$ &  $83.7314_{-0.0440}^{+0.0427}$
&  $7.9896_{-0.0048}^{+0.0038}$ &  $0.127582_{-0.000037}^{+0.000034}$ &
\citet{Mislis2010} \\ 
$ 414$ & $4980.467765_{-0.001178}^{+0.001166}$ & 
$84.0745_{-0.0726}^{+0.0579}$ &  $7.9897_{-0.0045}^{+0.0039}$ & 
$0.127585_{-0.000034}^{+0.000036}$ & \citet{Mislis2010} \\ 
$ 414$ &
$4980.467504_{-0.001389}^{+0.001041}$ &  $83.7421_{-0.0723}^{+0.0780}$ & 
$7.9900_{-0.0044}^{+0.0044}$ &  $0.127585_{-0.000037}^{+0.000035}$ &
\citet{Mislis2010} \\ 
$ 421$ & $4997.762881_{-0.000820}^{+0.000642}$ & 
$84.1880_{-0.0616}^{+0.0659}$ &  $7.9900_{-0.0040}^{+0.0046}$ & 
$0.127587_{-0.000033}^{+0.000037}$ & \citet{Scuderi2010} \\
\hline
\multicolumn{6}{c}{\textit{Kepler} data} \\
\hline
$ 404$ & $4955.763245_{-0.000044}^{+0.000050}$ &  $83.9673_{-0.0066}^{+0.0047}$&  $7.9883_{-0.0044}^{+0.0044}$ &  $0.127577_{-0.000034}^{+0.000037}$ \\ 
$ 405$ & $4958.233917_{-0.000048}^{+0.000047}$ & $83.9703_{-0.0056}^{+0.0055}$ &  $7.9886_{-0.0041}^{+0.0044}$ &  $0.127579_{-0.000032}^{+0.000036}$ \\ 
$ 406$ & $4960.704517_{-0.000054}^{+0.000042}$ &  $83.9662_{-0.0055}^{+0.0052}$ &  $7.9886_{-0.0041}^{+0.0039}$ &  $0.127575_{-0.000034}^{+0.000034}$ \\ 
$ 407$ & $4963.175144_{-0.000056}^{+0.000040}$ &  $83.9723_{-0.0056}^{+0.0063}$ &  $7.9897_{-0.0047}^{+0.0042}$ &  $0.127577_{-0.000036}^{+0.000031}$ \\ 
$ 408$ & $4965.645664_{-0.000045}^{+0.000048}$ &  $83.9784_{-0.0056}^{+0.0056}$ &  $7.9901_{-0.0040}^{+0.0046}$ &  $0.127578_{-0.000036}^{+0.000037}$ \\ 
$ 409$ & $4968.116325_{-0.000044}^{+0.000048}$ &  $83.9826_{-0.0052}^{+0.0062}$ &  $7.9900_{-0.0039}^{+0.0046}$ &  $0.127580_{-0.000033}^{+0.000037}$ \\ 
$ 410$ & $4970.586959_{-0.000051}^{+0.000043}$ &  $83.9672_{-0.0053}^{+0.0056}$ &  $7.9899_{-0.0043}^{+0.0040}$ &  $0.127587_{-0.000031}^{+0.000038}$ \\ 
$ 411$ & $4973.057564_{-0.000043}^{+0.000052}$ &  $83.9739_{-0.0062}^{+0.0049}$ &  $7.9876_{-0.0036}^{+0.0047}$ &  $0.127570_{-0.000032}^{+0.000039}$ \\ 
$ 412$ & $4975.528240_{-0.000048}^{+0.000043}$ &  $83.9795_{-0.0056}^{+0.0052}$ &  $7.9891_{-0.0043}^{+0.0037}$ &  $0.127578_{-0.000034}^{+0.000036}$ \\ 
$ 413$ & $4977.998789_{-0.000048}^{+0.000047}$ &  $83.9818_{-0.0056}^{+0.0058}$ &  $7.9904_{-0.0036}^{+0.0049}$ &  $0.127589_{-0.000034}^{+0.000037}$ \\ 
$ 414$ & $4980.469352_{-0.000051}^{+0.000045}$ &  $83.9698_{-0.0055}^{+0.0060}$ &  $7.9893_{-0.0044}^{+0.0041}$ &  $0.127570_{-0.000035}^{+0.000032}$ \\ 
$ 415$ & $4982.939951_{-0.000047}^{+0.000047}$ &  $83.9765_{-0.0051}^{+0.0061}$ &  $7.9874_{-0.0045}^{+0.0037}$ &  $0.127569_{-0.000031}^{+0.000038}$ \\ 
$ 416$ & $4985.410635_{-0.000047}^{+0.000047}$ &  $83.9681_{-0.0057}^{+0.0056}$ &  $7.9900_{-0.0048}^{+0.0037}$ &  $0.127581_{-0.000033}^{+0.000037}$ \\ 
$ 417$ & $4987.881204_{-0.000048}^{+0.000046}$ &  $83.9740_{-0.0056}^{+0.0055}$ &  $7.9904_{-0.0044}^{+0.0038}$ &  $0.127585_{-0.000033}^{+0.000036}$ \\ 
$ 418$ & $4990.351791_{-0.000048}^{+0.000048}$ &  $83.9804_{-0.0060}^{+0.0053}$ &  $7.9896_{-0.0043}^{+0.0042}$ &  $0.127580_{-0.000031}^{+0.000037}$ \\ 
$ 419$ & $4992.822527_{-0.000052}^{+0.000042}$ &  $83.9843_{-0.0062}^{+0.0054}$ &  $7.9896_{-0.0038}^{+0.0048}$ &  $0.127576_{-0.000035}^{+0.000036}$ \\ 
$ 420$ & $4995.293008_{-0.000046}^{+0.000044}$ &  $83.9716_{-0.0057}^{+0.0051}$ &  $7.9890_{-0.0040}^{+0.0041}$ &  $0.127578_{-0.000032}^{+0.000035}$ \\ 
$ 421$ & $4997.763604_{-0.000045}^{+0.000045}$ &  $83.9805_{-0.0062}^{+0.0051}$ &  $7.9895_{-0.0043}^{+0.0039}$ &  $0.127578_{-0.000032}^{+0.000038}$ \\ 
$ 424$ & $5005.175613_{-0.000044}^{+0.000049}$ &  $83.9741_{-0.0049}^{+0.0063}$ &  $7.9900_{-0.0041}^{+0.0041}$ &  $0.127589_{-0.000036}^{+0.000033}$ \\ 
$ 425$ & $5007.646198_{-0.000049}^{+0.000045}$ &  $83.9818_{-0.0056}^{+0.0053}$ &  $7.9884_{-0.0041}^{+0.0043}$ &  $0.127577_{-0.000032}^{+0.000037}$ \\ 
$ 426$ & $5010.116754_{-0.000047}^{+0.000047}$ &  $83.9715_{-0.0056}^{+0.0055}$ &  $7.9904_{-0.0039}^{+0.0043}$ &  $0.127583_{-0.000029}^{+0.000040}$ \\ 
$ 427$ & $5012.587380_{-0.000050}^{+0.000045}$ &  $83.9809_{-0.0059}^{+0.0056}$ &  $7.9899_{-0.0041}^{+0.0045}$ &  $0.127583_{-0.000038}^{+0.000030}$ \\ 
$ 430$ & $5019.999187_{-0.000052}^{+0.000042}$ &  $83.9850_{-0.0060}^{+0.0055}$ &  $7.9895_{-0.0045}^{+0.0048}$ &  $0.127589_{-0.000036}^{+0.000032}$ \\ 
$ 431$ & $5022.469822_{-0.000044}^{+0.000051}$ &  $83.9700_{-0.0058}^{+0.0052}$ &  $7.9901_{-0.0044}^{+0.0039}$ &  $0.127589_{-0.000033}^{+0.000037}$ \\ 
$ 432$ & $5024.940477_{-0.000044}^{+0.000050}$ &  $83.9771_{-0.0067}^{+0.0051}$ &  $7.9892_{-0.0046}^{+0.0043}$ &  $0.127581_{-0.000030}^{+0.000040}$ \\ 
$ 433$ & $5027.411097_{-0.000047}^{+0.000046}$ &  $83.9862_{-0.0060}^{+0.0050}$ &  $7.9897_{-0.0041}^{+0.0041}$ &  $0.127579_{-0.000036}^{+0.000035}$ \\ 
$ 434$ & $5029.881709_{-0.000048}^{+0.000043}$ &  $83.9768_{-0.0055}^{+0.0054}$ &  $7.9887_{-0.0043}^{+0.0040}$ &  $0.127577_{-0.000037}^{+0.000034}$ \\ 
$ 435$ & $5032.352219_{-0.000045}^{+0.000047}$ &  $83.9744_{-0.0058}^{+0.0057}$ &  $7.9899_{-0.0044}^{+0.0041}$ &  $0.127580_{-0.000032}^{+0.000038}$ \\ 
$ 436$ & $5034.822785_{-0.000045}^{+0.000047}$ &  $83.9830_{-0.0058}^{+0.0052}$ &  $7.9906_{-0.0045}^{+0.0039}$ &  $0.127587_{-0.000035}^{+0.000032}$ \\ 
$ 437$ & $5037.293526_{-0.000047}^{+0.000042}$ &  $83.9665_{-0.0060}^{+0.0052}$ &  $7.9893_{-0.0042}^{+0.0041}$ &  $0.127577_{-0.000036}^{+0.000034}$ \\ 
$ 438$ & $5039.764149_{-0.000049}^{+0.000046}$ &  $83.9742_{-0.0056}^{+0.0057}$ &  $7.9890_{-0.0041}^{+0.0044}$ &  $0.127582_{-0.000038}^{+0.000031}$ \\ 
$ 439$ & $5042.234769_{-0.000047}^{+0.000049}$ &  $83.9819_{-0.0063}^{+0.0054}$ &  $7.9916_{-0.0046}^{+0.0042}$ &  $0.127606_{-0.000036}^{+0.000035}$ \\ 
$ 440$ & $5044.705462_{-0.000046}^{+0.000049}$ &  $83.9731_{-0.0054}^{+0.0061}$ &  $7.9897_{-0.0044}^{+0.0041}$ &  $0.127580_{-0.000034}^{+0.000037}$ \\ 
$ 441$ & $5047.175958_{-0.000047}^{+0.000043}$ &  $83.9781_{-0.0054}^{+0.0059}$ &  $7.9889_{-0.0047}^{+0.0038}$ &  $0.127583_{-0.000035}^{+0.000034}$ \\ 
$ 442$ & $5049.646518_{-0.000050}^{+0.000046}$ &  $83.9691_{-0.0050}^{+0.0057}$ &  $7.9903_{-0.0042}^{+0.0039}$ &  $0.127591_{-0.000032}^{+0.000038}$ \\ 
$ 443$ & $5052.117194_{-0.000046}^{+0.000045}$ &  $83.9774_{-0.0047}^{+0.0063}$ &  $7.9902_{-0.0047}^{+0.0035}$ &  $0.127588_{-0.000038}^{+0.000031}$ \\ 
$ 444$ & $5054.587747_{-0.000046}^{+0.000051}$ &  $83.9775_{-0.0057}^{+0.0058}$ &  $7.9885_{-0.0042}^{+0.0044}$ &  $0.127581_{-0.000030}^{+0.000040}$ \\ 
$ 445$ & $5057.058368_{-0.000049}^{+0.000046}$ &  $83.9673_{-0.0054}^{+0.0055}$ &  $7.9890_{-0.0039}^{+0.0044}$ &  $0.127581_{-0.000034}^{+0.000037}$ \\ 
$ 446$ & $5059.529155_{-0.000049}^{+0.000044}$ &  $83.9787_{-0.0060}^{+0.0056}$ &  $7.9890_{-0.0045}^{+0.0042}$ &  $0.127580_{-0.000033}^{+0.000036}$ \\ 
$ 447$ & $5061.999681_{-0.000045}^{+0.000048}$ &  $83.9769_{-0.0057}^{+0.0054}$ &  $7.9895_{-0.0042}^{+0.0043}$ &  $0.127576_{-0.000035}^{+0.000036}$ \\ 
$ 449$ & $5066.940882_{-0.000045}^{+0.000048}$ &  $83.9786_{-0.0060}^{+0.0050}$ &  $7.9902_{-0.0041}^{+0.0043}$ &  $0.127589_{-0.000036}^{+0.000034}$ \\ 
$ 450$ & $5069.411456_{-0.000050}^{+0.000046}$ &  $83.9735_{-0.0058}^{+0.0050}$ &  $7.9896_{-0.0041}^{+0.0042}$ &  $0.127583_{-0.000036}^{+0.000033}$ \\ 
$ 451$ & $5071.882092_{-0.000044}^{+0.000045}$ &  $83.9814_{-0.0061}^{+0.0051}$ &  $7.9900_{-0.0038}^{+0.0048}$ &  $0.127587_{-0.000035}^{+0.000033}$ \\ 
$ 452$ & $5074.352766_{-0.000048}^{+0.000044}$ &  $83.9856_{-0.0054}^{+0.0058}$ &  $7.9910_{-0.0046}^{+0.0040}$ &  $0.127591_{-0.000033}^{+0.000034}$ \\ 
$ 453$ & $5076.823347_{-0.000048}^{+0.000047}$ &  $83.9805_{-0.0050}^{+0.0058}$ &  $7.9896_{-0.0035}^{+0.0045}$ &  $0.127578_{-0.000032}^{+0.000037}$ \\ 
$ 454$ & $5079.293970_{-0.000043}^{+0.000049}$ &  $83.9786_{-0.0058}^{+0.0051}$ &  $7.9904_{-0.0042}^{+0.0042}$ &  $0.127581_{-0.000029}^{+0.000039}$ \\ 
$ 455$ & $5081.764470_{-0.000054}^{+0.000043}$ &  $83.9759_{-0.0059}^{+0.0051}$ &  $7.9892_{-0.0038}^{+0.0044}$ &  $0.127581_{-0.000035}^{+0.000035}$ \\ 
$ 456$ & $5084.235245_{-0.000048}^{+0.000045}$ &  $83.9985_{-0.0048}^{+0.0058}$ &  $7.9907_{-0.0043}^{+0.0037}$ &  $0.127594_{-0.000032}^{+0.000036}$ \\ 
$ 457$ & $5086.705793_{-0.000044}^{+0.000048}$ &  $83.9832_{-0.0061}^{+0.0053}$ &  $7.9902_{-0.0040}^{+0.0045}$ &  $0.127581_{-0.000034}^{+0.000035}$ \\ 
$ 460$ & $5094.117546_{-0.000047}^{+0.000046}$ &  $83.9744_{-0.0046}^{+0.0065}$ &  $7.9914_{-0.0043}^{+0.0041}$ &  $0.127594_{-0.000033}^{+0.000037}$ \\ 
$ 461$ & $5096.588269_{-0.000049}^{+0.000042}$ &  $83.9794_{-0.0052}^{+0.0057}$ &  $7.9904_{-0.0039}^{+0.0042}$ &  $0.127585_{-0.000038}^{+0.000032}$ \\ 
$ 462$ & $5099.058744_{-0.000046}^{+0.000049}$ &  $83.9875_{-0.0057}^{+0.0059}$ &  $7.9908_{-0.0048}^{+0.0042}$ &  $0.127593_{-0.000034}^{+0.000037}$ \\ 
$ 463$ & $5101.529370_{-0.000044}^{+0.000051}$ &  $83.9845_{-0.0051}^{+0.0058}$ &  $7.9903_{-0.0043}^{+0.0043}$ &  $0.127599_{-0.000032}^{+0.000037}$ \\ 
$ 464$ & $5103.999990_{-0.000045}^{+0.000048}$ &  $83.9797_{-0.0058}^{+0.0050}$ &  $7.9895_{-0.0041}^{+0.0043}$ &  $0.127592_{-0.000033}^{+0.000037}$ \\ 
$ 465$ & $5106.470706_{-0.000043}^{+0.000049}$ &  $83.9855_{-0.0061}^{+0.0049}$ &  $7.9899_{-0.0041}^{+0.0044}$ &  $0.127586_{-0.000035}^{+0.000035}$ \\ 
$ 466$ & $5108.941275_{-0.000044}^{+0.000046}$ &  $83.9943_{-0.0061}^{+0.0056}$ &  $7.9897_{-0.0037}^{+0.0048}$ &  $0.127581_{-0.000030}^{+0.000038}$ \\ 
$ 467$ & $5111.411898_{-0.000048}^{+0.000042}$ &  $83.9798_{-0.0059}^{+0.0051}$ &  $7.9896_{-0.0044}^{+0.0041}$ &  $0.127585_{-0.000037}^{+0.000035}$ \\ 
$ 469$ & $5116.353078_{-0.000048}^{+0.000042}$ &  $83.9814_{-0.0049}^{+0.0062}$ &  $7.9907_{-0.0039}^{+0.0046}$ &  $0.127591_{-0.000033}^{+0.000035}$ \\ 
$ 470$ & $5118.823862_{-0.000049}^{+0.000046}$ &  $84.0001_{-0.0051}^{+0.0058}$ &  $7.9922_{-0.0038}^{+0.0047}$ &  $0.127599_{-0.000038}^{+0.000033}$ \\ 
$ 471$ & $5121.294291_{-0.000046}^{+0.000047}$ &  $83.9872_{-0.0056}^{+0.0053}$ &  $7.9899_{-0.0046}^{+0.0040}$ &  $0.127590_{-0.000036}^{+0.000032}$ \\ 
$ 473$ & $5126.235583_{-0.000047}^{+0.000044}$ &  $83.9888_{-0.0061}^{+0.0052}$ &  $7.9893_{-0.0040}^{+0.0043}$ &  $0.127586_{-0.000037}^{+0.000035}$ \\ 
$ 474$ & $5128.706192_{-0.000046}^{+0.000046}$ &  $83.9941_{-0.0058}^{+0.0051}$ &  $7.9910_{-0.0036}^{+0.0044}$ &  $0.127586_{-0.000033}^{+0.000037}$ \\ 
$ 475$ & $5131.176834_{-0.000047}^{+0.000042}$ &  $83.9822_{-0.0045}^{+0.0061}$ &  $7.9911_{-0.0039}^{+0.0040}$ &  $0.127592_{-0.000035}^{+0.000034}$ \\ 
$ 476$ & $5133.647423_{-0.000048}^{+0.000042}$ &  $83.9931_{-0.0058}^{+0.0052}$ &  $7.9909_{-0.0041}^{+0.0041}$ &  $0.127593_{-0.000035}^{+0.000033}$ \\ 
$ 477$ & $5136.118086_{-0.000052}^{+0.000043}$ &  $83.9800_{-0.0059}^{+0.0055}$ &  $7.9907_{-0.0046}^{+0.0042}$ &  $0.127592_{-0.000043}^{+0.000029}$ \\ 
$ 478$ & $5138.588677_{-0.000047}^{+0.000048}$ &  $83.9806_{-0.0060}^{+0.0055}$ &  $7.9928_{-0.0049}^{+0.0040}$ &  $0.127603_{-0.000035}^{+0.000035}$ \\ 
$ 479$ & $5141.059254_{-0.000049}^{+0.000044}$ &  $83.9840_{-0.0050}^{+0.0060}$ &  $7.9905_{-0.0043}^{+0.0041}$ &  $0.127588_{-0.000031}^{+0.000038}$ \\ 
$ 480$ & $5143.529887_{-0.000043}^{+0.000045}$ &  $83.9797_{-0.0049}^{+0.0061}$ &  $7.9888_{-0.0041}^{+0.0043}$ &  $0.127577_{-0.000034}^{+0.000036}$ \\ 
$ 481$ & $5146.000644_{-0.000046}^{+0.000042}$ &  $83.9659_{-0.0059}^{+0.0053}$ &  $7.9895_{-0.0047}^{+0.0038}$ &  $0.127580_{-0.000036}^{+0.000033}$ \\ 
$ 482$ & $5148.471133_{-0.000046}^{+0.000046}$ &  $83.9813_{-0.0057}^{+0.0052}$ &  $7.9898_{-0.0043}^{+0.0043}$ &  $0.127584_{-0.000033}^{+0.000036}$ \\ 
$ 483$ & $5150.941691_{-0.000047}^{+0.000043}$ &  $83.9799_{-0.0055}^{+0.0054}$ &  $7.9899_{-0.0039}^{+0.0042}$ &  $0.127591_{-0.000037}^{+0.000034}$ \\ 
$ 484$ & $5153.412285_{-0.000045}^{+0.000044}$ &  $83.9755_{-0.0062}^{+0.0049}$ &  $7.9902_{-0.0042}^{+0.0040}$ &  $0.127583_{-0.000036}^{+0.000035}$ \\ 
$ 486$ & $5158.353590_{-0.000049}^{+0.000045}$ &  $83.9865_{-0.0053}^{+0.0057}$ &  $7.9909_{-0.0040}^{+0.0043}$ &  $0.127592_{-0.000034}^{+0.000034}$ \\ 
$ 487$ & $5160.824096_{-0.000043}^{+0.000046}$ &  $83.9753_{-0.0062}^{+0.0054}$ &  $7.9909_{-0.0040}^{+0.0047}$ &  $0.127592_{-0.000035}^{+0.000035}$ \\ 
$ 488$ & $5163.294713_{-0.000051}^{+0.000041}$ &  $83.9746_{-0.0054}^{+0.0056}$ &  $7.9907_{-0.0042}^{+0.0042}$ &  $0.127587_{-0.000037}^{+0.000032}$ \\ 
$ 489$ & $5165.765446_{-0.000044}^{+0.000051}$ &  $83.9836_{-0.0053}^{+0.0058}$ &  $7.9897_{-0.0036}^{+0.0047}$ &  $0.127587_{-0.000035}^{+0.000032}$ \\ 
$ 490$ & $5168.236001_{-0.000047}^{+0.000042}$ &  $83.9756_{-0.0058}^{+0.0056}$ &  $7.9913_{-0.0045}^{+0.0041}$ &  $0.127593_{-0.000036}^{+0.000031}$ \\ 
$ 491$ & $5170.706625_{-0.000043}^{+0.000045}$ &  $83.9790_{-0.0061}^{+0.0048}$ &  $7.9905_{-0.0038}^{+0.0046}$ &  $0.127590_{-0.000032}^{+0.000036}$ \\ 
$ 492$ & $5173.177189_{-0.000043}^{+0.000049}$ &  $83.9814_{-0.0052}^{+0.0057}$ &  $7.9907_{-0.0044}^{+0.0039}$ &  $0.127598_{-0.000035}^{+0.000034}$ \\ 
$ 493$ & $5175.647822_{-0.000051}^{+0.000043}$ &  $83.9678_{-0.0059}^{+0.0050}$ &  $7.9901_{-0.0039}^{+0.0042}$ &  $0.127581_{-0.000036}^{+0.000033}$ \\ 
$ 494$ & $5178.118439_{-0.000048}^{+0.000039}$ &  $83.9781_{-0.0055}^{+0.0060}$ &  $7.9907_{-0.0043}^{+0.0048}$ &  $0.127598_{-0.000036}^{+0.000034}$ \\ 
$ 495$ & $5180.589204_{-0.000053}^{+0.000040}$ &  $83.9747_{-0.0057}^{+0.0048}$ &  $7.9892_{-0.0037}^{+0.0043}$ &  $0.127582_{-0.000036}^{+0.000035}$ \\ 
\hline 
\end{longtable}
}

\end{appendix}

\end{document}